\documentclass[12pt]{article}
\usepackage[margin=1in]{geometry}                 
\usepackage{setspace}
\usepackage{graphicx}
\usepackage{amsmath}
\usepackage{amssymb}
\usepackage{bm}
\usepackage{url}
\usepackage{epstopdf}
\usepackage{times}
\usepackage{courier}
\usepackage{authblk}

\usepackage{algorithmic}
\usepackage{textcomp}
\usepackage{xcolor}
\usepackage{subcaption}
\usepackage{diagbox}
\usepackage{tabulary}
\usepackage{mathtools}
\usepackage{listings, multicol}
\title{WHO-IS: Wireless Hetnet Optimization \\ using Impact Selection}
\author{Thomas Sandholm, Irene Macaluso, Sayandev Mukherjee}
\affil{CableLabs}

\begin{document}
\maketitle

\begin{abstract}
We propose a method
to first identify users
who have the most negative impact on the
overall network performance, and then offload them
to an orthogonal channel.
The feasibility of such
an approach is verified using real-world traces,
network simulations, and a lab experiment that employs
multi-homed wireless stations.
In our experiment, as offload target, 
we employ LiFi IR transceivers, and as the primary network
we consider a typical Enterprise Wi-Fi setup.
We found that a limited number of users can impact
the overall experience of the Wi-Fi network negatively,
 hence motivating targeted offloading. In our simulations
and experiments we saw that the proposed solution can improve
the collision probability with 82\% and achieve a 61 percentage point
air utilization improvement compared to random offloading, 
respectively.
\end{abstract}

\section{Introduction}\label{sec:introduction}
With the emergence of wireless IoT devices,
an increased emphasis on remote video conferencing, and an ever increasing demand
from new bandwidth-hungry applications, such as AR/VR, Wi-Fi
networks are struggling to keep up. New spectrum availability and more spectrum efficient
protocols mitigate congestion but do not fully solve the problem, due to the trade-offs involved in 
operating on different spectrum bands, e.g. range, throughput trade-offs. In an office environment 
this trend is exacerbated with dense AP and station deployments, 
where limited orthogonal Wi-Fi 
channels force use of narrower bands and hence lower throughput
capacity to avoid excessive interference. In such environments, saturation of airtime utilization,
and spiking collision probabilities cause packet delays that lead to 
a poor user experience, especially for latency-sensitive applications. 

These trends have led to renewed interest in meeting the demand with heterogeneous networks (HetNets) composed of
technologies and protocols operating in
non-interfering spectrum bands.

One such technology often proposed for load balancing, and capacity enhancement is Light Fidelity (LiFi).  The first generation of LiFi systems operated in the visible light spectrum.  However, newer generations of LiFi systems use the infrared (IR) spectrum.
With the introduction of IR-based as opposed to visible light communication 
LiFi no longer suffers from many of the initial drawbacks 
that made it impractical for general use, such as light dimming degradation and poor uplink performance due to glare.
Furthermore, throughput performance of commercially available hardware is starting to match
that of Wi-Fi. 
The current downsides of LiFi include expensive hardware, 
limited reach (line of sight, LoS), and intrusive deployments
with essentially a dedicated LiFi antenna for each receiver. LiFi APs can handle tens of clients concurrently.  
However, the very limited range and sensitivity to 
receiving orientation angle (ROA) make the 1-1 mapping of antenna to receiver the most
practical setup at the present time, at least until the multi-user time-slicing standards 
currently in development are finalized. LiFi APs typically operate on the same spectrum band, 
so inter-cell interference becomes an issue if deployments are too dense.

To address some of these practical problems we utilize a setup that involves a LiFi antenna
that can be steered through a pan-tilt servo mount to serve one dedicated user over LiFi among a group
of users within range. Beyond cost savings, there are many benefits to directing the antenna more precisely to the receiver including achieving longer distances, allowing a much larger area to be covered while mitigating inter-cell interference, and avoiding multi-user access degradation. In our experimental setup a single antenna can easily serve a cluster of four standard-sized office
cubicles while mounted in a ceiling 3-4m high whereas the LiFi hardware only allows a 68\% angle
reach within about 2 meters, limiting it to transmit from something like an office lamp to a device
on a desk. Although the technology will undoubtedly improve both in reach and angle coverage
we believe the servo steering technique could be feasible to extend the range of LiFi while still
providing dedicated Line of Sight (LoS) communication, which is considered to be a security benefit of LiFi.

Our primary contribution in this work is a predictive model, based on
neural networks, that: 
\begin{itemize}
\item \emph{predicts} which Wi-Fi station should be offloaded
to an orthogonal channel, in our case LiFi, 
\item \emph{given} current measurements both from the 
primary and offloading target network 
\item \emph{to optimize} network KPIs, 
such as air-time utilization and collision probability.
\end{itemize}

We validate our work with real-world trace analysis, Wi-Fi network simulations (in NS3),
and a lab experiment with commercial-grade Wi-Fi and LiFi hardware and off-the-shelf end-user
devices. The remainder of the paper is organized as follows. 

We discuss related work in Section~\ref{sec:relatedwork}, and motivate
our approach using an analysis of public Wi-Fi traces in Section~\ref{sec:traceanalysis}.
In Section~\ref{sec:model} we define our problem more formally, followed by an evaluation
using simulations (Section~\ref{sec:simulation}) and experiments (Sections~\ref{sec:system}
and~\ref{sec:experiment}). Finally, we provide concluding remarks in Section~\ref{sec:conclusion}.

\section{Related Work}\label{sec:relatedwork}
Predicting or selecting one or more users to offload from a given network or access point to another based on some optimality criterion has been studied for a long time under the topics of ``user association'' and ``user handoff.'' For example,~\cite{5462204} applied a game theoretic framework to analyze association in a network with High Speed Downlink Packet Access (HSDPA) and LTE, while ~\cite{7248853} formulated a stochastic game to model non-cooperative users competing for limited resources from multiple cellular base stations. A general utility-optimizing formulation for user association in a heterogeneous network was proposed in~\cite{7848549}, while~\cite{7024778} applied a multiple attribute decision making method with careful selection of user attributes to reduce computational complexity. Treating user association as a combinatorial optimization problem instead, a stochastic decision framework was proposed and analyzed in~\cite{6883721}. A fuzzy logic approach to designing handoff between a WLAN and a cellular network to reduce call dropping probability was studied in~\cite{1628378, 4677973}, while~\cite{7151085, 8611376} applied a constrained Markov Decision Process formulation instead, and ~\cite{8003266} proposed a user association scheme based on load-balancing using a cell-breathing scheme.

To the best of our knowledge,
our work is the first approach of using 
an auto-trained Neural Network\footnote{we refer to our approach as a neural network as it may be implemented as a deep neural network (DNN) with many hidden layers in some deployments, but simpler more shallow networks in others.} (NN) to predict the
user that is optimal to offload (e.g. from Wi-Fi to LiFi) using a KPI impact perspective.
The approach has, however, been inspired by previous
contributions in the areas of Machine Learning (ML) and Wi-Fi/LiFi HetNets.

\subsection{ML-driven HetNets}
Interference is classical problem in HetNets, and in ~\cite{anand2022machine}
the authors propose an ML classification and offloading scheme to improve
co-tier interference between femtocells. Support Vector Machine (SVM), Random Forest (RF), Artificial
Neural Networks (ANN) and Convolutional Neural Networks (CNN) are all evaluated as
alternative models for interference classification and subsequent offloading decisions. 
CNN and RF were the top performers. Interference is similar to our collision
probability KPI; however, our solution differs in the way we formulate the offloading
decision, in that we try to predict the user who impacts the current network
most negatively.

A reinforcement learning solution is proposed for joint power control and user association in a millimeter wave heterogeneous network in~\cite{8692747}. In~\cite{7997424}, a recommender model was proposed to map users to access points (LTE or Wi-Fi), while in~\cite{7925955}, the authors modeled the user association problem as a restless multi-armed bandit and exploited individual user behavior characteristics to maximize long-term expected system throughput.
In \cite{xiong2013clustering} $k$-means clustering is used to classify users to
improve handover decisions across HetNets based on user context.
In \cite{lei2017deep} a DNN is trained offline to make optimal cache placement
decisions in a HetNet. 

Our approach differs from all of the above in that we can accommodate multiple KPIs into our optimization through our problem formulation based on the negative impact score.

\subsection{Wi-Fi/LiFi HetNets}
From the early days of LiFi there has been work considering how to best
manage a hybrid Wi-Fi and LiFi HetNet network~\cite{haas2015lifi,wang2016fuzzy,wu2017joint,wang2017load,ayyash2016coexistence}.
These studies were evaluated with custom simulations, and assumed fixed LiFi beam directions. They also focused on
improving user satisfaction and throughput as opposed to network KPIs as in our study. Furthermore, the models proposed
were either using linear programming models or optimization heuristics based on game theory or genetic algorithms. Based on
the complexity of Wi-Fi alone, we believe machine learning (ML) approaches such as neural networks (NNs) are a more promising
basis for a solution and would also scale better to more users.

More recently, this idea has been revisited~\cite{vijayaraghavan2021delay} to formulate a resource allocation optimization problem
minimizing delay and meeting a minimum data rate by assigning resource shares across Wi-Fi and LiFi APs. 
In that work, the LiFi antennas are deployed and directed statically to cover 
an entire meeting room. Since they allow multiple users on the same AP, they need to account for interference both on the LiFi
and the Wi-Fi bands. Furthermore, since the LiFi beams are not targeted, the best signal of an AP is not guaranteed to be where
the user is located who receives the signal. We believe that ML techniques such as NNs are better suited than 
traditional optimization problem formulations in capturing the complexity of both Wi-Fi
and LiFi networks, and we think the allocations can be more efficient with dedicated LiFi channels since the beam range
is so limited. Given the current cost of a LiFi antenna it is also both a cost issue and deployment hassle to litter 
the ceiling with one antenna for each position a device may be located. Furthermore, most office ceilings are higher than
the 2m range of current LiFi transmitters.

In \cite{bechadergue2020industrial} beam forming inspired by mmWave technology is mentioned as a future direction of LiFi,
and the general problem of spectrum shortage is highlighted as a future research problem where LiFi offloading could help.
We also note that, according to this overview which is based on the hardware we are using, current chipsets do not support
handover between Wi-Fi and LiFi and thus mobility is a problem. However, the new converged 802.11bb specification
does support handover, and once that is implemented in chipsets it becomes more interesting to study which
users to offload rather than making the switch more efficient, which helps explain the focus of our work. 

\section{Motivation}\label{sec:traceanalysis}
We analyze packet capture traces from Wi-Fi deployments to:
\begin{itemize}
\item motivate our general approach of defining and selecting so-called \emph{negative impact
users} to offload to LiFi (see below), 
\item evaluate candidate statistics as impact predictors, and
\item validate some impact prediction models.
\end{itemize}

\subsection{Negative Impact Score}
First, we need to define what we mean by negative impact
in order to identify Wi-Fi users (STAs) that are candidates
for offloading to LiFi.

The trace is divided into equally-sized time segments indexed by $t=1,2,\dots$, and we then measure $r_t$,
the overall packet retry probability across all captured packets
in each time segment $t$. 

We also track all users who started sending
or receiving packets (entered) or stopped sending or receiving packets (dropped out)
in each time segment $t$. 

Suppose a new user $u$ entered the system in time segment $t$, but all other active users in time segment $t-1$ remained active in time segment $t$ and no other users entered or departed the system in time segment $t$.  (This is likely to be the case if the time segments are short in duration and the system is not very heavily loaded.)  Then we say that user $u$ had a \emph{negative impact} on the system if the overall packet retry probability $r_t$ in $t$, the first time segment \emph{with} user $u$ active, is \emph{greater} than the overall packet retry probability $r_{t-1}$ in the previous time segment:
\begin{equation}
	r_t > r_{t-1},
\end{equation}
or equivalently,
\begin{equation}
	\Delta^{\mathrm{e}}_{u,t} \equiv r_t - r_{t-1} > 0.
	\label{eq:Delta_e}
\end{equation}
Note that since $r_{t-1}$ and $r_t$ are both aggregate system measurements, so is their difference $\Delta_t = r_t - r_{t-1}$.  However, because of our assumption that the only change to the system between time intervals $t-1$ and $t$ is the entry of the user $u$, we can attribute the change in overall packet retry probability to $u$, hence we are justified in attaching the subscript $u$ to $\Delta_t$.  The superscript $\mathrm{e}$ denotes the \emph{entry} of this user $u$ into the system in time segment $t$.

Similarly, suppose that user $u'$ was active in time segment $t'-1$ and departed the system in time segment $t'$, while all other active users in time segment $t'-1$ remained active in time segment $t'$ and no other users entered or departed the system in time segment $t'$.  Then we say that user $u'$ had a \emph{negative impact} on the system if the overall packet retry probability $r_{t'}$ in $t'$, the first segment \emph{without} user $u'$ active, is \emph{lower} than the overall packet retry probability $r_{t'-1}$ in the previous time segment:
\begin{equation}
	r_{t'} < r_{t'-1},
\end{equation}
or equivalently,
\begin{equation}
	\Delta^{\mathrm{d}}_{u',t'} \equiv r_{t'} - r_{t'-1} < 0,
	\label{eq:Delta_d}
\end{equation}
where again the subscript $u'$ for the aggregate system measurement $\Delta_{t'}$ is justified because of our assumptions above, and the superscript $\mathrm{d}$ denotes the \emph{departure} of user $u'$ from the system.

Now consider a single user $u$, and suppose that in the trace, $u$ is seen to enter the system in time segments $t_1, t_2, \dots, t_n$ and depart the system in time segments $t'_1, t'_2, \dots, t'_m$.  As before, we assume that the duration of each time segment is short enough that in each of these time segments, $u$ is the only user to either enter or depart the system, and all other users retain their state of activity or inactivity unchanged from the immediately prior time segment. We also assume that the total time for the trace is short enough for us to assume (quasi)-stationarity, so that we may model $\Delta^{\mathrm{e}}_{u,t_1},\dots,\Delta^{\mathrm{e}}_{u,t_n}$ as independent identically distributed (i.i.d.) random variables with common expected value $\mu^{\mathrm{e}}(u)$, and similarly model $\Delta^{\mathrm{d}}_{u,t'_1},\dots,\Delta^{\mathrm{d}}_{u,t'_m}$ as i.i.d.~random variables with common expected value $\mu^{\mathrm{d}}(u)$.  Note that from~\eqref{eq:Delta_e} and~\eqref{eq:Delta_d}, it follows that $\mu^{\mathrm{e}}(u)>0$ and $\mu^{\mathrm{d}}(u)<0$ respectively.  The mean magnitudes $|\mu^{\mathrm{e}}(u)| = \mu^{\mathrm{e}}(u)$ and $|\mu^{\mathrm{d}}(u)| = -\mu^{\mathrm{d}}(u)$ may be seen as measures of the negative impact of user $u$ entering and departing the system respectively.
We can then define the Negative Impact Score (NIS) of user $u$ as the sum of the above two negative impact measures of $u$ entering and departing the system:
\begin{equation}
	\mathrm{NIS}(u) = |\mu^{\mathrm{e}}(u)| + |\mu^{\mathrm{d}}(u)| = \mu^{\mathrm{e}}(u) - \mu^{\mathrm{d}}(u).
\end{equation}
In practice, the two expectations $\mu^{\mathrm{e}}(u)$ and $\mu^{\mathrm{d}}(u)$ are estimated by
\begin{equation}
	\hat{\mu}^{\mathrm{e}}(u) = \frac{\Delta^{\mathrm{e}}_{u,t_1} + \cdots + \Delta^{\mathrm{e}}_{u,t_n}}{n}
\end{equation}
and
\begin{equation}
	\hat{\mu}^{\mathrm{d}}(u) = \frac{\Delta^{\mathrm{d}}_{u,t'_1} + \cdots + \Delta^{\mathrm{d}}_{u,t'_m}}{m}
\end{equation}
respectively.

%

\subsection{STA Measurements}
\label{sec:STA_meas}
Since many users may enter or drop out in the same segments, 
an accurate negative impact score for a single user relies on sampling over many segments
where users enter and drop out many times. Our metric here can thus be seen
as an approximation of measuring the impact directly on a per-user basis, which
is not possible in this case due to the fact that we use public data sets without
this granularity. However, in a real system deployment it may 
be possible to predict $\mathrm{NIS}$ from STA statistics measured more
directly. To determine the feasibility of different statistics
we collect the following measurements from the traces for each STA:
\begin{itemize}
\item {\it rx } received bytes per second. 
\item {\it tx } sent bytes per second.
\item {\it size } packet size (bytes).
\item {\it rssi } RSSI signal (dBm).
\item {\it phyrate } PHY rate based on MCS obtained (Mbps).
\item {\it packets } number of packets.
\item {\it iat } inter-arrival time of packets (s).
\item {\it retries } retry probability of packets.
\end{itemize}

\subsection{Data Sets}
We use a public data set captured in five different
venues around Portland State Univerity, Oregon (PSU)~\cite{pdxdata}, as well as our own
private radio capture in an Enterprise office setting (ENT).
The private capture was necessary to obtain {\it rssi}
and {\it phyrate} measurements, as well as to do a longitudinal
study over an extended period of time (12 hours).

\subsection{Impact Outliers}
Given our approach of selecting individual STAs to offload
to LiFi, we want to verify whether there are a few users (more than one, but not too many) with high enough negative
impact scores to make it:
\begin{enumerate}
\item worthwhile to offload individual users to improve the
overall network performance significantly, and
\item non-trivial to make the optimum selection of the user(s) to offload to LiFi.
\end{enumerate}

\begin{figure}
        \centering
        \begin{subfigure}[b]{0.49\textwidth}
                \includegraphics[width=\textwidth]{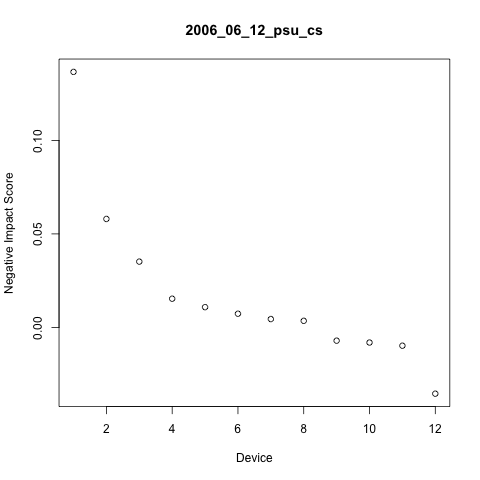}
        \end{subfigure}
        \begin{subfigure}[b]{0.49\textwidth}
                \includegraphics[width=\textwidth]{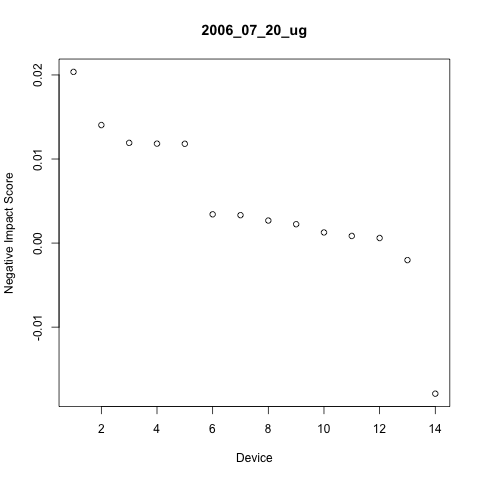}
        \end{subfigure}
        \begin{subfigure}[b]{0.49\textwidth}
                \includegraphics[width=\textwidth]{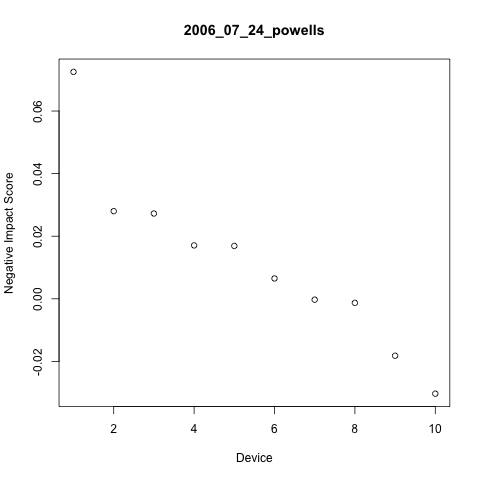}
        \end{subfigure}
        \begin{subfigure}[b]{0.49\textwidth}
                \includegraphics[width=\textwidth]{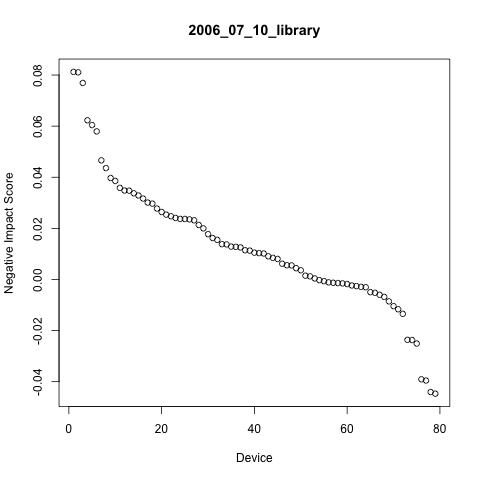}
        \end{subfigure}
        \begin{subfigure}[b]{0.49\textwidth}
                \includegraphics[width=\textwidth]{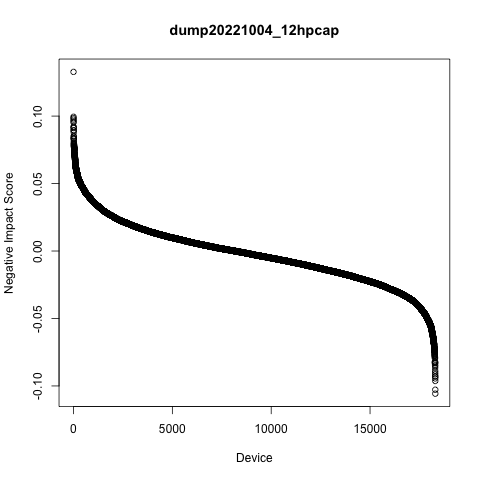}
        \end{subfigure}
        \caption{Ranked NIS scores}
        \label{nisrank}
\end{figure}

To determine whether there are outliers in terms of high negative
impact we rank, and plot negative impact scores across STAs using
the different data sets. Note that a higher $\mathrm{NIS}$ means that
the STA has a more negative impact on the overall network,
and the $x$-axis is the NIS rank of a particular user,
so we are looking for outliers in the top left of the plots.

From Fig.~\ref{nisrank} we see that all data sets revealed high NIS
outliers. We should note that for some data set the top
outlier was further away from the average than others
as can be seen from the scale of the $y$-axis. Note that
the NIS numbers are in probability units as NIS is computed
from collision probabilities.

\subsection{Metric Predictor Analysis}
Next, we study the ability to predict NIS scores from
the STA measurements listed in Sec.~\ref{sec:STA_meas}. We use ANOVA
analyses of measurements statistics and look at F-score
significance as a measure of which statistics
show promise in predicting NIS for different data sets.
The ANOVA analysis results can be seen in Table~\ref{T:anova}.

\begin{table}[htbp]
	\caption{Significant metrics (5\% level) in ANOVA analysis for different data sets.}
\begin{center}
\begin{tabular}{|l|l|l|}
\hline
	\textbf{Data Set} & \textbf{Top Metric (Lowest F score)} & \textbf{Significant Metrics} \\
\hline
	PSU CS & rx ($1.7\times10^{-1}$) & - \\
\hline
	PSU UG &  size ($4.3\times10^{-4}$) & size\\
\hline
	PSU Library & size ($1.2\times10^{-5}$) & size, tx\\
\hline
	PSU Powells &  iat ($8.8\times10^{-2}$) & - \\
\hline
	ENT & size ($2\times10^{-16}$) & size, rx, rssi, iat \\
\hline
\end{tabular}
\label{T:anova}
\end{center}
\end{table}

\subsection{Impact Class Prediction}
Based on the ANOVA analysis we now take the metric
deemed as the best predictor in terms of lowest F-score (see Table~\ref{T:anova})
and predict the NIS with that predictor for a
random user and train with the other users.
We then pick 100 different random users and compute 
the average prediction success rate (over these 100 predictions). 

Instead of trying to predict the NIS of a user directly, we simplify the prediction problem
by trying to predict only a binary value representing the tercile of this user's NIS value:
$0$ if this user's NIS is in the top $1/3$ of all users' NIS values, and $1$ otherwise.
We also compare our model predictor
with a random predictor that picks 0 with 
probability $1/3$ and 1 with probability $2/3$. 
We make this prediction in 30 rounds and compute the average
and standard deviation of the success rates, shown in Table~\ref{T:classpredict}.

\begin{table}[htbp]
	\caption{NIS Class Prediction Mean Success Rate $\pm1.96\sigma$.}
\begin{center}
\begin{tabular}{|l|l|l|}
\hline
	\textbf{Data Set} & \textbf{Linear Regression} & \textbf{Random} \\
\hline
	PSU CS & $.70\pm.04$ & $.54\pm.05$ \\
\hline
	PSU UG & $.72\pm.03$ & $.55\pm.05$ \\
\hline
	PSU Library & $.72\pm.04$ & $.55\pm.04$ \\
\hline
	PSU Powells &  $.67\pm.04$ & $.52\pm.04$ \\
\hline
	ENT & $.74\pm.04$ & $.58\pm.04$ \\
\hline
\end{tabular}
\label{T:classpredict}
\end{center}
\end{table}

In summary we have shown with this trace analysis
that there is an opportunity to predict outlier users
with a high negative impact on the overall health of
the network, and that STA measurements can predict
this impact.

\section{Model}\label{sec:model}

Each LiFi AP is enhanced with a pan-tilt unit that can orientate the AP to cover different areas. We assume a finite number $C_l$ of spatial configurations for LiFi AP $l$, with spatial configuration $i$ corresponding to \emph{user area} (the service area for that LiFi AP configuration $i$) $u_{i, l}$, $i=1,\dots,C_l$. For example, in an office environment a user area could correspond to a cubicle. A LiFi AP \emph{coverage area} is the union of all the user areas that can be served by the AP. We will assume that a user area can be served by at most one LiFi AP, i.e., the LiFi coverage areas do not overlap. It is worth noting that this decomposition is not an exact geometric representation of the environment.

Assuming that it is possible to collect a set of measurements for each device in the network at regular intervals on both Wi-Fi and LiFi, we denote by $\bm{w}_{n}=[w_{n,1}(t),\ldots,w_{n,K_{\mathrm{Wi}}}(t)] $
and $\bm{l}_{n}=[l_{n,1}(t),\ldots,l_{n,K_{\mathrm{Li}}}(t)]$ respectively the set of $K_{\mathrm{Wi}}$ Wi-Fi measurements and the set of $K_{\mathrm{Li}}$ LiFi measurements for device $n$ at time $t$.

For each user area $u\in\{u_{1,l},u_{2,l},\ldots,u_{C_l,l}\}$ of LiFi AP $l$, we can aggregate\footnote{The aggregation of measurements may be implemented in different ways, e.g. using the sum, mean, max, or min.} the Wi-Fi and LiFi measurements of all the devices in that user area and denote the aggregated measurements as $\bm{w}_u^{(a)}$ and $\bm{l}_u^{(a)}$ respectively.   

An important point to note is that we do not want ``ping-ponging,'' i.e., frequent transfer of an STA between Wi-Fi and LiFi, or frequent switching of the STA served by a given LiFi AP.  This is similar to the ping-ponging problem of handover in a mobile cellular wireless network, and the remedy is the same, namely the use of a hysteresis factor to retain the association of an STA with a LiFi AP for a certain interval of time after the STA has been offloaded to LiFi.  This hysteresis may be implemented in several ways; in the present work, we implement it by ``boosting'' the $\bm{l}_u^{(a)}$ measurements by the hysteresis factor $h_u^{\mathrm{LiFi}} > 1$, which in general may be dependent on the user area $u$.  In other words, the vector 
\begin{equation}
\bm{E}_l(t)=[\bm{w}_{u_{1,l}}^{(a)}, h_{u_{1,l}}^{\mathrm{LiFi}}\bm{l}_{u_{1,l}}^{(a)}, \bm{w}_{u_{2,l}}^{(a)}, h_{u_{2,l}}^{\mathrm{LiFi}}\bm{l}_{u_{2,l}}^{(a)}, \ldots, \bm{w}_{u_{C_l,l}}^{(a)}, h_{u_{C_l,l}}^{\mathrm{LiFi}}\bm{l}_{u_{C_l,l}}^{(a)}]
\label{eq:context_E}
\end{equation}
contains the  Wi-Fi and LiFi measurements of all the devices that can connect to LiFi AP $l$, aggregated per user area.  

We model each LiFi AP as an autonomous agent that can decide which device(s) should be selected to be served\footnote{Multiple devices in the same user area could be selected or a single device could be targeted. In the latter case, a one-to-one mapping between user areas and devices need to be established.} by LiFi and change its orientation accordingly. Each LiFi AP makes a decision with the goal of optimizing the overall network performance $f$, which in general could be defined as a scalar function of several KPIs in the combined network\footnote{For example, the user throughput, retransmissions, collisions, or air utilization can all be incorporated in the definition of $f$.}. To do this, each LiFi AP learns a mapping between the current network state $\bm{E}(t)$, the possible actions, and the overall network performance. LiFi AP $l$ can use as network state the vector $\bm{E}_l(t)$, i.e. consider only the measurements corresponding to all the user areas in its coverage area, or it can include also additional measurements, for example the measurement vectors $\bm{E}_j(t)$ of nearby LiFi APs.

We can formulate the problem as a \emph{Contextual Multi-Armed Bandit} (CMAB), where the network state $\bm{E}(t)$ is the context. The mapping between context, actions, and the resulting network performance may take different forms. It may be modeled as a function from the (context, action) pair to the resulting network performance. Another option is to model the mapping from context to the network performance of each action. In both cases, the mapping can be learned by a neural network.

\section{Simulation}\label{sec:simulation}
In the trace analysis we were able to show the opportunity and
ability to predict the negative impact score of users to select 
candidates for offloading to improve the overall
network performance for all users. 

\subsection{Layout and geometry}
Since the analysis was done on static traces, we have no
way of measuring the actual impact of offloading these users.
Therefore, we now simulate an offloading scenario using NS3, where we model a network with only a
single Wi-Fi AP containing eight users: (i) a cluster of four STAs that are candidates
to be offloaded, and (ii) another cluster of four STAs that serve
as background users impacting the performance. 

In this simulation setup, there is no LiFi AP; instead, the effect of offloading an STA from Wi-Fi to LiFi is simulated by simply dropping that STA from the simulation, thereby allowing us to simulate only the Wi-Fi network on each NS3 run.

The four STAs that are candidates for offloading to LiFi are closer to the 
(Wi-Fi) AP and have \emph{average 
throughput} 20\% lower than the four other STAs that serve as background users. Although it may be counter-intuitive to have the candidate users have \emph{lower} average throughput than the background users, it has the effect that any outlier (in terms of traffic) amongst the candidate users therefore has an outsize impact on the system KPIs when it drops out.  At the same time, this setup allows for ``headroom'' for the traffic at this outlier candidate, ensuring that with high probability, even the outlier traffic does not hit (and get capped at) the maximum throughput possible in the system.

\subsection{Workload trace generation}
We replay workloads with a generative adversarial network (GAN)-based synthetic workload generation tool, MASS~\cite{sandholm2021mass}, trained on a public data set from Telefonica~\cite{telefonicadata}.

On each NS3 run, MASS is used to generate a 100-epoch long trace for each of our 8 STAs. 
The traces are split into multiple sections, each section of duration 10 epochs, which we call a {\it period}. The upload and download
rate can vary for each STA from one period to the next. In other words, each trace may be seen as a \emph{time series} of (upload, download) rate pairs for a particular STA, with as many such pairs as there are periods in the trace.  Moreover, the upload and/or download rate for that STA can only change at the boundary of a period.

For each STA, each epoch of each period of the trace constitutes a MASS-generated workload with the appropriate (uplink/downlink) rate for that period, replayed for 2~{\!s}. Since the client iPerf processes (on the STAs) and the LiFi offload controller process (on the LiFi AP) are not synchronized, a STA may be selected for offload to LiFi at any time, and the offloading will take effect from the next epoch in the trace of that STA.  The short 2-second duration of each epoch therefore ensures that the maximum delay in offloading an STA to LiFi is 2~{\!s}.

The maximum requested TCP download and upload rates are both set to 100~{\!Mbps},
and a 20~{\!MHz} wide 5~{\!GHz} 802.11ac channel is used for all STAs and the AP.

\begin{figure}
        \centering
                \includegraphics[width=\textwidth]{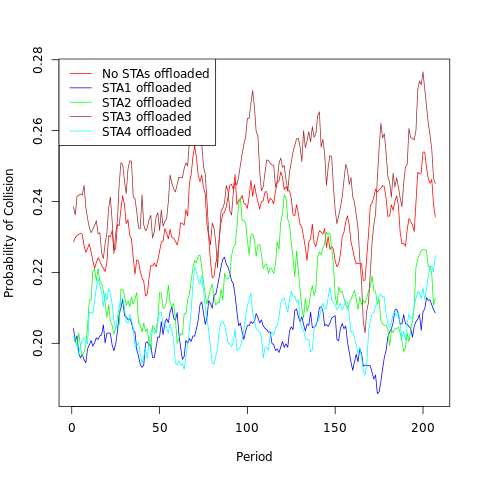}
        \caption{Simulation Collision Time Series}
        \label{simcol}
\end{figure}

\subsection{Statistics collected}
We then run five simulations (in NS3) for each 10-epoch period in the trace, corresponding to the scenarios where no candidate STA is offloaded, and each of the four candidate STAs is
offloaded respectively.  For each of these five simulations in each period, we collect the pcap Wi-Fi packet trace between each STA and the AP.  This allows us to capture all
upload and download packets, as well as retries and other statistics such as effective send and receive rates\footnote{The effective upload and download rates are obtained by smoothing using an average over 5 periods.},
and packet inter-arrival times. We call the resulting statistics our {\it measurements}. 

For each NS3 run, we also collect a global
statistic of average retry probability across all packets sent in the system. We call this statistic the {\it collision} KPI,
and it is our measure of overall network performance. The goal is to reduce this collision KPI by selecting the best STA to be offloaded
(among the four candidates).  

Fig.~\ref{simcol} shows the collision KPI for all five decisions (i.e., no STA offloaded or one of the possible four candidate STAs offloaded) during one simulation. It is clear that it is not optimal to select just one of these STAs and offload it alone throughout the entire duration of the trace.

We perform 50 NS3 runs, i.e., we repeat the steps of generating a 100-epoch (= 10-period) trace, running simulations (five per period, as described above) and collecting statistics, 50 times.

\subsection{Predicting offload candidates}
To evaluate different offloading predictors, we first train the prediction models with the STA measurements across all 8 STAs
based on which state the system is in, i.e. no STA offloaded or, one of the four candidate STAs offloaded. During training, the prediction model output is set to the collision KPI in the next period for each of these five possible states. This setup
allows us to use the model to predict the KPI in the next period, given any possible state in the current period.

\subsubsection{Prediction model training and inference}
The prediction models are trained with the data from the first 8 periods of the (10-period-long) traces from the 50 NS3 runs (with five simulations per period, one per state), resulting in a training set with $8\times5\times50=2000$ samples (\textit{\#periods $\times$ \#states $\times$ \#simulations}). 

The data of the remaining 2 periods of each trace are used for testing, for a total of $2\times 50 = 100$ predictions (\textit{\#periods $\times$ \#traces}). These predictions correspond to the selection of a user for offload in each of the final 2 periods of each of the 50 traces.

When predicting the best user to offload in the next period, we always assume that all four candidate STAs are present in the Wi-Fi network, i.e., the present state of the system is one where no STA has been offloaded. It should be noted that each of the 100 test predictions requires four predicted {\it collision} KPIs to be computed, one for each of the candidate STAs. The best user to offload is chosen as the STA that, when offloaded, results in the minimum collision KPI as predicted by the prediction model.

\subsubsection{Prediction model evaluation}
From our NS3 runs, we observed that the maximum collision probability reduction for a genie-aided (clairvoyant) predictor, which always predicts the STA that minimizes collision KPI upon being offloaded, 
is about $16$\%. The {\it collision improvement score}, $\mathrm{cis}$, of a predictor $p$ is defined as:
\begin{equation}
	\mathrm{cis} = \frac{P^c_0-P^c_p}{P^c_0-P^c_{\mathrm{cv}}}
\end{equation}
where $P^c_0$ is the collision probability when no STA is offloaded, $P^c_p$ is the
collision probability when the STA picked by $p$ is offloaded\footnote{Actually, $p$ predicts the collision KPIs when each of the four STAs is offloaded, so $P^c_p$ is the smallest of these four collision KPIs predicted by $p$.},
and $P^c_{\mathrm{cv}}$ is the collision probability for the clairvoyant predictor. A clairvoyant predictor would hence have a collision improvement score of 
$100$\%. 

The prediction accuracy of predictor $p$ simply measures how many times it picked the correct STA to offload, i.e., how many times $P^c_p = P^c_{\mathrm{cv}}$. For a baseline, we also define the so-called \emph{random predictor}, which does not attempt to predict the collision KPI upon the offloading of any of the four STAs, but instead simply directly chooses one of the four STAs at random (with the same probability $1/4$ for each STA) for offloading. Since this randomly-selected STA is expected to match the STA chosen by the clairvoyant predictor only 25\% of the time, this random predictor should be expected to have a prediction accuracy of 25\%.

\subsubsection{Comparing predictors}
We compare the performance of the following predictors, listed below in descending order of sophistication:
\begin{itemize}
\item[NN]: A Neural Network model;
\item[LR]: A Linear Regression model; 
\item[COL]: A na\"{i}ve model that predicts that the collision KPIs corresponding to the offloading of each of the four candidate STAs in the \emph{next} period simply equal the corresponding observed collision KPIs from the simulations in the \emph{current} period;
\item[RAND]: The baseline random predictor described above, which simply selects one of the four candidate STAs at random with the same probability of $1/4$ for offloading.
\end{itemize}
Note that even the random
predictor is expected to improve the collision rate on average, as offloading any user should reduce the traffic
on the channel under contention. 

Table~\ref{T:simpredict_5roll} shows the comparison between the different predictors. While the prediction accuracy of all models except RAND are similar, NN outperforms all the other models in terms of collision improvement score. Indeed, the prediction accuracy is not necessarily a good indicator of a predictor's performance. For example, RAND has the lowest prediction accuracy but significantly outperforms COL in terms of reducing the collisions in the network, as measured by $\mathbf{cis}$. This happens because often, even if a predictor fails to predict the best STA to offload, the selected user is still impacting the KPI significantly. For example, if we compare the NN and LR performance, we notice that their prediction accuracy is quite close. However, NN selects STAs whose offloading improves the network performance more than that of the STAs selected by LR. In summary, we see
an 82 percent improvement in collision probability when using the NN offloader compared to only 31 percent with the
random model.  

\begin{table}[htbp]
	\caption{Simulation Prediction Accuracy and Collision Improvement Scores ($\pm\mathit{SE}$).}
\begin{center}
\begin{tabular}{|l|l|l|}
\hline
	\textbf{Predictor} & \textbf{Prediction Accuracy  } & $\mathbf{cis}$ \\
\hline
	NN & $.51\pm.05$ & $.82\pm.1$ \\
\hline
	LR & $.49\pm.05$ & $.72\pm.13$ \\
\hline
	COL & $.50\pm.05$ & $.31\pm.28$ \\
\hline
	RAND & $.24\pm.04$ & $.47\pm.29$ \\
\hline
\end{tabular}
\label{T:simpredict_5roll}
\end{center}
\end{table}

\begin{figure}
        \centering
                \includegraphics[width=\textwidth]{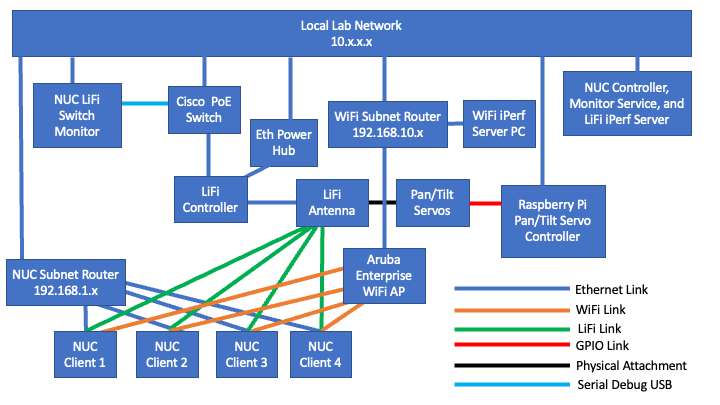}
	\caption{Physical Network Links. It should be noted that only one of the LiFi links (shown in green) is present at any given time.}
        \label{linkarch}
\end{figure}

\begin{figure}
        \centering
                \includegraphics[width=\textwidth]{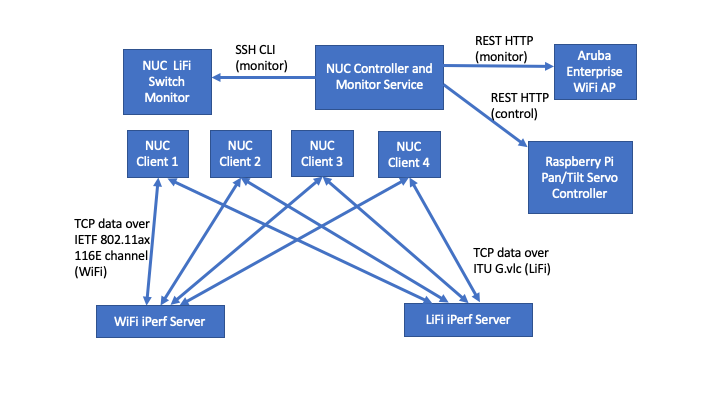}
	\caption{Communication Paths and Protocols.}
        \label{commarch}
\end{figure}

\begin{figure}
        \centering
                \includegraphics[width=10cm]{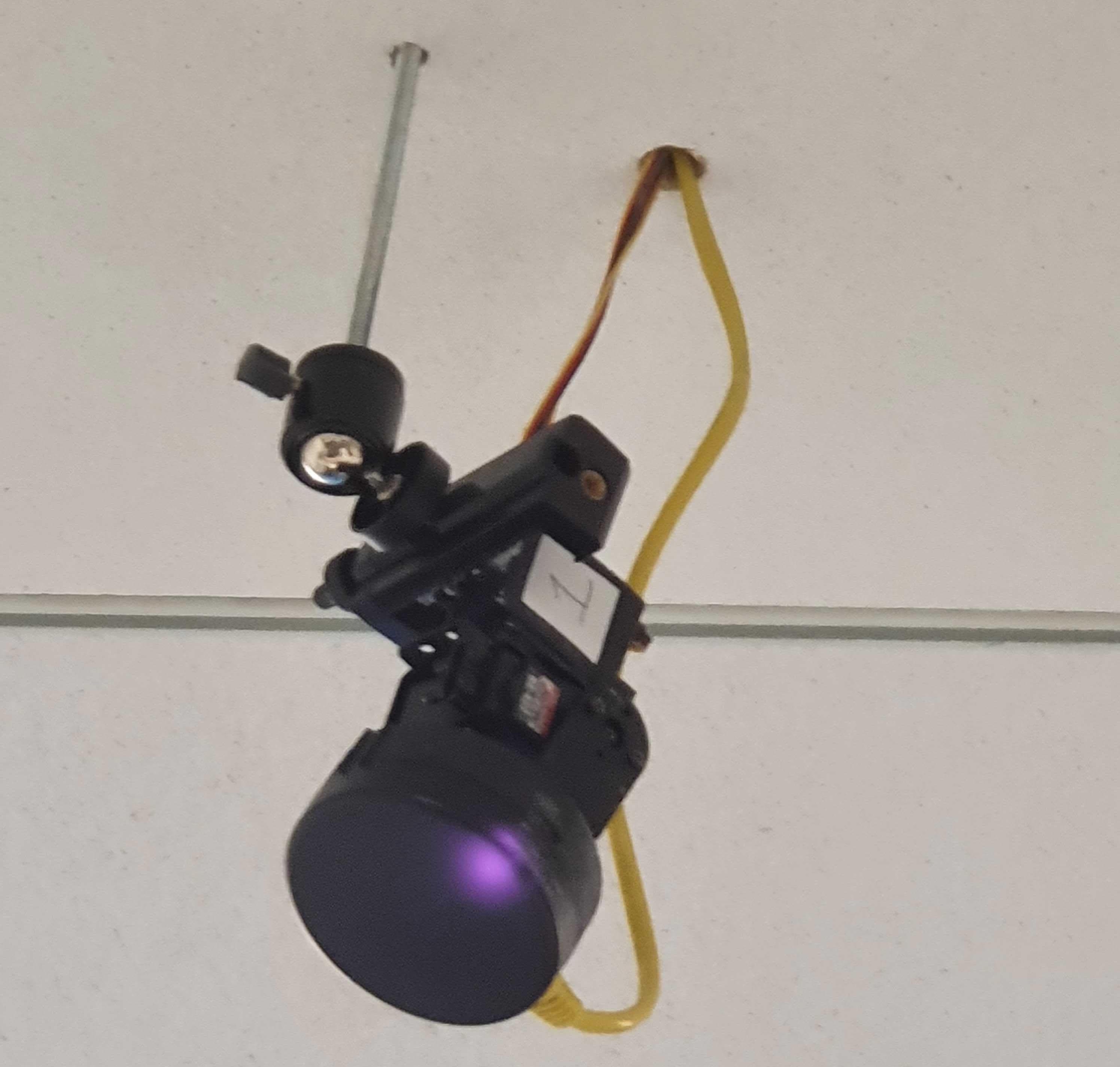}
        \caption{LiFi Antenna mounted on Pan-Tilt servos and connected to RaspberryPi in ceiling.}
        \label{lifiantenna}
\end{figure}

\begin{figure}
        \centering
                \includegraphics[width=16cm]{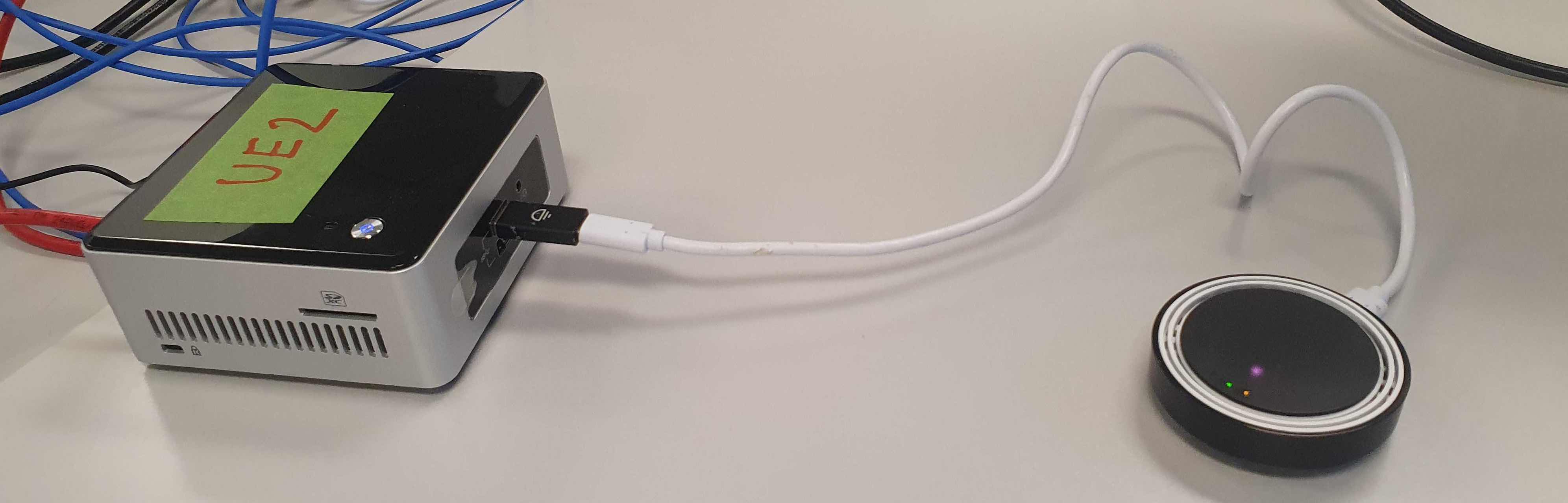}
        \caption{NUC with LiFi receptor dongle.}
        \label{lificlient}
\end{figure}

\begin{figure}
        \centering
                \includegraphics[width=\textwidth]{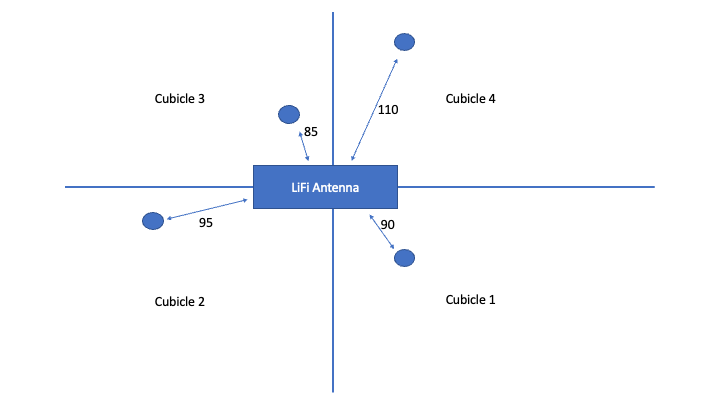}
	\caption{Experiment Setup. All distances denote shortest path from sender to receiver in inches.}
        \label{explab}
\end{figure}

\begin{figure}
        \centering
                \includegraphics[width=16cm]{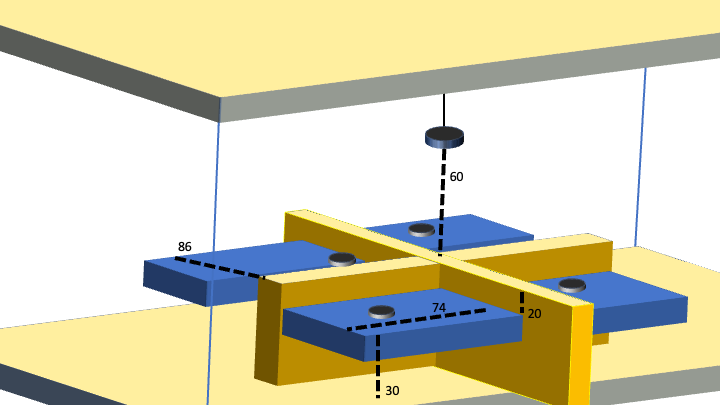}
	\caption{Experiment Setup, 3D View. All distances denote inches.}
        \label{explab3d}
\end{figure}

\section{Experimental Setup}\label{sec:system}
The purpose of our experimental system is to verify and reproduce the simulation
results with real hardware radios and optical links in a lab setting that mimics an Enterprise Wi-Fi offloading use case.
A key difference between simulations and experiments is also the introduction of a LiFi beam steering
mechanism designed to reuse a LiFi antenna over many clients positioned in a greater area, in our case a 4-person
cubicle grid. One LiFi/Wi-Fi multi-homed wireless station is placed in each cubicle.

The antenna is directed in such a way that only one wireless station or client gets 
dedicated LiFi connectivity at any time. All
clients always have Wi-Fi to fall back on, but continuously probe and connect to LiFi if it is available. Our goal here is to improve the Wi-Fi KPI by picking the best client to offload to LiFi
at any given time, based on Wi-Fi and LiFi measurements and KPI predictions.

We use LiFi hardware from Oledcomm (LiFiMax), and Wi-Fi hardware from Aruba (550 Series).
To monitor the LiFi traffic we also do switch port monitoring with a Cisco switch
connected to the LiFi controller, which in turn receives all LiFi traffic from the
LiFi antenna.

We developed an antenna steering
REST API service on a Raspberry Pi, which
is connected via GPIO pins to servos controlling the angle of the LiFi antenna beams
with pulse-width modulation signals. For mechanical antenna control, we use a Lynxmotion Pan and Tilt Kit with two HiTec HS-422 
180 degree servo motors.

We built central offloading control and monitoring services in Python that implement the NN prediction model using
Tensorlow and collect
KPI and measurements from the Wi-Fi and LiFi traffic in the network. 
The offloading control, LiFi/Wi-Fi multi-homed clients, and the switch monitor are all deployed on
Intel NUCs running Ubuntu 22.04 LTS with the 5.15 kernel. 

The physical link architecture is depicted in Fig.~\ref{linkarch}
and the communication architecture in Fig.~\ref{commarch}.

Photos of tbe LiFi AP, mounted in the ceiling, and one of the four NUC LiFi clients are shown
in Fig.~\ref{lifiantenna} and Fig.~\ref{lificlient}.

The LiFi client and antenna position layout is visualized in
Figs.~\ref{explab} and~\ref{explab3d}. The four LiFi beam positions
are fixed and calibrated before the experiment and the LiFi client dongles remain stationary.

To simplify communication over different network interfaces, LiFi and Wi-Fi for experiment traffic and
Wired for experiment orchestration, we created two subnets so that each interface gets an IP in 
a different subnet.

\begin{figure}
        \centering
                \includegraphics[width=16cm]{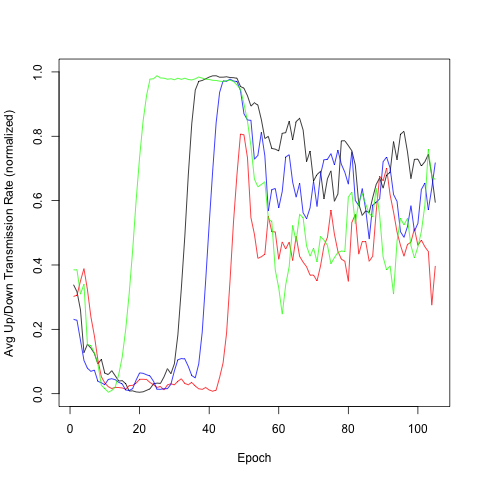}
	\caption{Workload trace used for the initial exploration stage (first quarter) and the predictions (last three quarters).}
        \label{exptrace}
\end{figure}

\begin{figure}
        \centering
                \includegraphics[width=\textwidth]{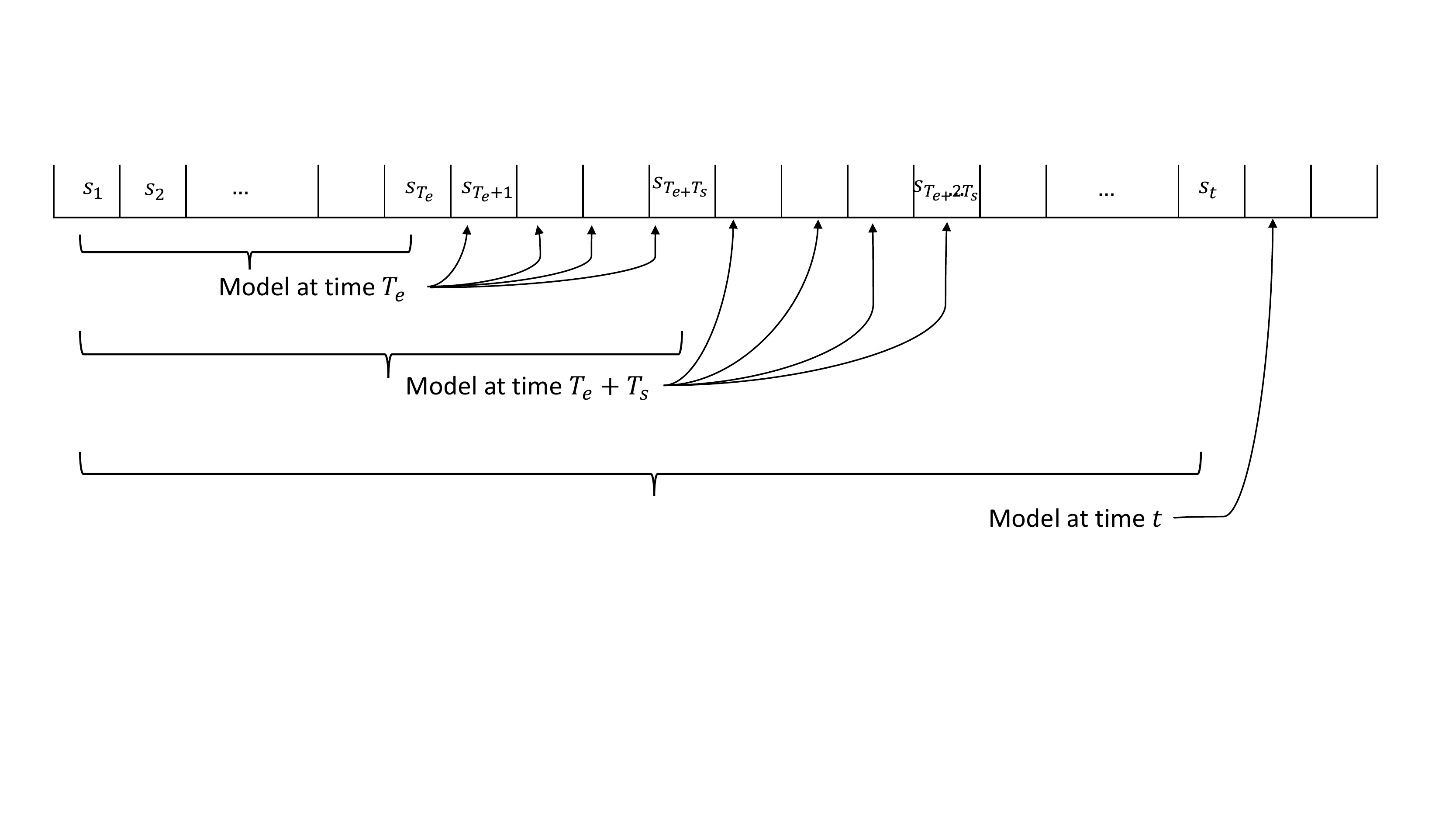}
	\caption{At the beginning of each experiment  samples $\{s_1,s_2,\ldots, s_{T_e}\}$ are collected and used to train the model at time $T_e$. This model is used to decide the LiFi antenna orientation that will be used in the next $T_s$ time units, where a time unit is the time required to collect a sample. The model is updated at time $T_e+T_s$, using the most recent $T_s$ samples. }
        \label{diagram-pred-training}
\end{figure}

\subsection{Traffic Replay}
The four wireless stations run iPerf3 clients against two sets of iPerf3
servers, one in the LiFi subnet and one in the Wi-Fi subnet\footnote{Owing to the design of iPerf, a single iPerf server cannot handle both upload and download sessions; instead, we need a ``set'' of two iPerf servers, one for upload and one for download, for each subnet.}. 
One client-server pair is used for upload and one for download, for a total of 8 streams per iPerf server set.

It is important to ensure that the load on the CPU of the server due to iPerf does not become the bottleneck on network performance.
In the LiFi case only at most two streams from a single station will be active, so the load due to iPerf is light, and the LiFi iPerf set is deployed on a NUC.
On the other hand, the Wi-Fi iPerf set may in some runs serve all four STAs concurrently, so its load on the server CPU may be high. So we run the Wi-Fi iPerf set on a more powerful CPU (MacBook Air).

Traffic is generated with a specific bitrate based on trace files as in the simulation. 
The traces (one per station) comprise a sequence of 100 epochs\footnote{A 5-period smoothing is applied, as in the simulations.},
and are replayed in each experiment condition. 

For each STA, each epoch of each period of the trace constitutes a MASS-generated workload with the appropriate (uplink/downlink) rate for that period, replayed for 10~{\!min}. The STAs use a wired connection (in order not to load the wireless network) to 
check for LiFi availability (with a curl to the LiFi server) every 10~{\!s}. Since the client processes (on the STAs) and the LiFi offload controller process (on the LiFi AP) are not synchronized, an STA may be selected for offload to LiFi at any time, and the offloading will take effect from the next epoch in the trace of that STA.  Forcing the STAs to check with the LiFi offload controller every 10~{\!s} ensures that the maximum delay in offloading an STA to LiFi is 10~{\!s}.

Fig.~\ref{exptrace} shows the total — upload and download — workload per epoch of the four STAs. 

\subsection{Offloading Prediction}
\label{sec:off_pred}
The offloading control service monitors Wi-Fi and LiFi measurements as well as Wi-Fi KPI every 30-40s. Each experiment takes almost 17 hours and corresponds to about 1500 collected samples, where each sample consists of Wi-Fi and LiFi measurements and the corresponding KPI. We run 5 independent experiments for each of the three prediction models — NN, LR, and RAND — for a total experiment run time
of about 250~{\!h}.

Each experiment starts with a initial exploration during which the LiFi antenna positions are selected in round-robin and $T_e=400$ samples (out of about 1500 total) are collected. This data is used to train the initial prediction model. This means that about the first quarter of the traces shown in Fig.~\ref{exptrace} are used for this initial stage.

During the entire experiment, including the initial exploration, switches between LiFi positions are only 
allowed every $T_s=4$ samples, or about every 2~{\!min}. After the initial exploration, a prediction model is trained and used to make decisions on which station should have the LiFi antenna directed to it at any given
time. Before making any new decision, the prediction model is updated with the $T_s=4$ most recent samples. Fig.~\ref{diagram-pred-training} depicts this process.

\section{Experiment}\label{sec:experiment}
\subsection{Defining the prediction models}
As explained in Sec.~\ref{sec:model}, we formulate the problem of deciding which device should be selected for LiFi as a CMAB. In this section we evaluate different models that approximate the mapping between the (context, action) pair in the CMAB and the resulting network performance.

First, we extract air utilization ($\mathrm{air}$) as the KPI from the Aruba Wi-Fi AP
on a scale from 0 to 100. Since it is
more natural to consider a higher KPI better we invert these measurements as follows:
\begin{equation}
  \mathrm{kpi} = 100 - \mathrm{air}
\end{equation}
The goal of the proposed system is to optimize future KPI in the overall hybrid Wi-Fi/LiFi network.
This is done in a two-step process: first, given the current network state $\bm{E}(t)$, a future KPI is estimated for each possible
action (position), and then the actions are sorted to retrieve
the top position. 

We pre-process the Wi-Fi and LiFi measurements and extract features that are then used as the context $\bm{E}(t)$ of the CMAB. The most significant difference between the inputs to the models in simulation and in the experiment is the inclusion of LiFi measurements in the latter. The models trained and tested in simulation did not include LiFi measurements since no LiFi module is currently available in NS3. The other difference is that in the case of the experiment the raw Wi-Fi and LiFi measurements are processed to extract features (the impact vector explained below) to be able to train the models with a small number of samples. 

\subsection{Inputs to the prediction models}

The input to the prediction model is the context of the CMAB, which from Sec.~\ref{sec:model} is defined by~\eqref{eq:context_E}
\[
\bm{E}(t)=[\bm{w}_{u_1}^{(a)}, h^{\mathrm{LiFi}}\bm{l}_{u_1}^{(a)}, \bm{w}_{u_2}^{(a)}, h^{\mathrm{LiFi}}\bm{l}_{u_2}^{(a)}, \ldots, \bm{w}_{u_C}^{(a)}, h^{\mathrm{LiFi}}\bm{l}_{u_C}^{(a)}], 
\]
where for brevity of notation, we have dropped the LiFi AP index $l$ and assumed the same hysteresis factor\footnote{Recall from Sec.~\ref{sec:model} that we want to avoid ping-ponging LiFi offload between different positions using a hysteresis factor.  Moreover, the Wi-Fi and LiFi statistics are measured differently and should be scaled appropriately in order to be comparable.  As suggested in Sec.~\ref{sec:model}, we accomplish both goals via
LiFi to Wi-Fi impact conversion by multiplying with a hysteresis factor $h_u^{\mathrm{LiFi}} > 1$, set to be the same for all LiFi user areas $u$ for simplicity.} $h^{\mathrm{LiFi}}$ for all $C$ user areas $u_1, \dots, u_C$.  In our experimental setup, $C=4$, corresponding to the four possible positions that the LiFi antenna can be pointed to.  In addition to $\bm{E}(t)$, our context for the CMAB also includes the present position that the LiFi antenna is pointed to.\footnote{This is represented by a 4-element one-hot indicator vector with a $1$ in the position that the antenna is pointed to. The indicator vector allows us to probe for KPI estimates for different LiFi antenna positions for a given context.}

Since exactly one STA (among four) can be served by the LiFi antenna in each of its four positions, we further simplify $\bm{E}(t)$ by replacing the Wi-Fi and LiFi measurements in the user areas $u_1,\dots,u_C$ by the corresponding quantities computed at the four STAs themselves.

The Wi-Fi measurements for the four STAs are their relative normalized traffic loads on the downlink and uplink respectively, as defined through the following Softmax operation for each direction (downlink or uplink) separately:
\begin{equation}
	\bm{w}^{\mathrm{dir}}(t) = \mathrm{softmax}\left(\frac{N_1^{\mathrm{dir}}(t) - N_{1,\min}^{\mathrm{dir}}}{N_{1,\max}^{\mathrm{dir}} - N_{1,\max}^{\mathrm{dir}}}, \dots, \frac{N_4^{\mathrm{dir}}(t) - N_{4,\min}^{\mathrm{dir}}}{N_{4,\max}^{\mathrm{dir}} - N_{4,\max}^{\mathrm{dir}}}\right),
\end{equation}
where $\mathrm{dir}$ denotes the direction (uplink or downlink), and $N_{n, \min}^{\mathrm{dir}}$, $N_{n, \max}^{\mathrm{dir}}$, and $N_n^{\mathrm{dir}}(t)$ are respectively the minimum, maximum, and instantaneous (at time $t$) traffic load generated by STA $n$ in direction $\mathrm{dir}$, for each $n=1,\dots,4$.  The softmax operation is designed to emphasize outliers, as the goal is to select the LiFi antenna position (or equivalently, the STA to be offloaded) yielding the highest performance as measured by the collision KPI.

The above vector $\bm{w}^{\mathrm{dir}}(t)$ is dense (fully populated) if no STA has been offloaded to LiFi.  If an STA is offloaded from Wi-Fi to LiFi, then its corresponding entry in the vector $\bm{w}^{\mathrm{dir}}(t)$ is set to 0 and the corresponding entry in the similarly-defined vector of LiFi measurements $\bm{l}^{\mathrm{dir}}(t)$, is calculated instead and scaled by the hysteresis factor $h^{\mathrm{LiFi}}$. We have empirically determined that
$h^{\mathrm{LiFi}} = 3.5$ gives good results not only for the experiments reported here, but also on other workloads and in different settings.

%
%
%
%

\subsection{Experimental measurements}
The sampled measurements are summarized in Table~\ref{T:experimentmeasurement}.

\begin{table}[htbp]
	\caption{Wi-Fi and LiFi Station Measurements.}
\begin{center}
\begin{tabular}{|l|l|}
\hline
{\bf Wi-Fi} & frames\_in\_fps \\
 & frames\_out\_fps \\
\hline
{\bf LiFi} & up\_throughput \\
 & down\_throughput \\
\hline
\end{tabular}
\label{T:experimentmeasurement}
\end{center}
\end{table}

The Wi-Fi measurements were collected with the Aruba REST API, and the LiFi measurements were collected using
the Cisco switch CLI (using {\it show interface summary}). We note that our models normalize all measurements,
so the fact that the units of different measurements are different has no impact on the predictions, and thus
were chosen based on what was easiest to measure.

\subsection{Evaluating the prediction models}
The performance of the three predictor models is evaluated starting from the end of the initial exploration stage to the end of the traces. Let this time duration span $T$ samples, which we label $k=1,2,\dots,T$.  As previously noted, each of the three models is evaluated on $E=5$ independent experiments. Let us denote by $f_{e,m}(k)$, $k=1,\dots,T$ the KPI values (airtime, in our case) observed over these samples in the experiment $e$ with prediction model $m$. 

For each model $m$, we define its normalized performance over these samples as: 
\begin{equation}
	r_{e,m}(k) = \frac{f_{e,m}(k) - \min_m f_{e,m}(k)}{\max_m f_{e,m}(k)-\min_m f_{e,m}(k)}, \quad k=1,\dots,T.
\end{equation}	

\begin{figure}
        \centering
                \includegraphics[width=16cm]{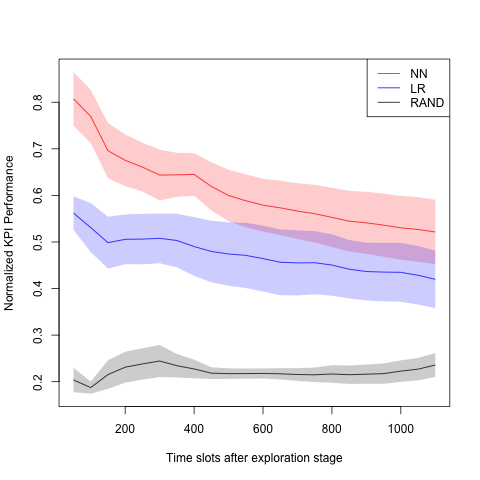}
	\caption{Plot of average normalized KPI performance $\overline{r}_m(\cdot)$ for each model $m$.}
        \label{expresult}
\end{figure}

Then, we compute the running-average normalized performance of $m$ over these samples as:
\begin{equation}
	[\overline{r}_{e,m}(1), \dots, \overline{r}_{e,m}(T)],
\end{equation}
where $\overline{r}_{e,m}(1) = r_{e,m}(1)$ and for $k=2,\dots,T$, the $k$th element in the above is the average of the previous $k-1$ normalized performance values: 
\begin{equation}
	\overline{r}_{e,m}(k) = \frac{1}{k-1}\sum_{j=1}^{k-1} \overline{r}_{e,m}(j), \quad k=2,\dots,T.
\end{equation}   
Finally, we compute the running-average normalized performance of each model $m$ averaged across the $E=5$ experiments as: 
\begin{equation}
 \overline{r}_m(k) = (1/E)\sum_{e=1}^E \overline{r}_{e,m}(k)
\end{equation} 
Fig.~\ref{expresult} plots $\overline{r}_m(k)$ versus $k$ for $k \geq 50$, for each of the three models $m$. Table~\ref{T:experimentresults} summarizes the results for the different models $150$ samples after the end of the initial exploration, which is 400 samples as reported in Sec.~\ref{sec:off_pred}. 


\begin{table}[htbp]
	\caption{Summary of Experiment Results after 550 samples from the beginning of the experiment.
	\textbf{CP} is collision probability; 
	\textbf{LiFi} is LiFi traffic multiplier compared to the optimal condition; 
	the KPI is the average airtime across the previous 150 samples, i.e. since the end of the initial exploration stage of 400 samples, and across all experiments $e\in\{1,2,\dots,5\}$.}

\begin{center}
\begin{tabular}{|l|c|c|l|}
\hline
\textbf{Model} & \textbf{KPI} (airtime) & \textbf{CP} (collision probability) & \textbf{LiFi} \\
\hline
NN & $.80$ & $.10$ & $1.4$ \\
LR & $.54$  & $.13$ & $1.1$ \\
RAND & $.19$ & $.19$ & $.16$ \\
Optimal & $1$ & $.11$ & $1$ \\
Worst & $0$ & $.19$ & $0.41$ \\
\hline
\end{tabular}
\label{T:experimentresults}
\end{center}
\end{table}


We note that the improvement of NN over LR is reduced the further away from training the predictions are, indicating
that re-calibration or random exploration could be motivated. 

In summary, we have seen that the 
NN model can improve the airtime utilization KPI compared to the random model with 61 percentage points (from 19\% to 80\%),
bring the collision probability, despite not being part of the optimization, from 19\% to 10\%, while offloading almost a factor of 10 more traffic over the LiFi link (.16 to 1.4), in the 150 sample prediction case.

\section{Discussion and Conclusion}\label{sec:conclusion}
Although our lab experiment made use of an iPerf client that
is capable of detecting whether LiFi is present one cannot expect
all applications to seamlessly take advantage of a new LiFi
network popping up. To this end we have experimented with a number
of other tools to help take advantage of both networks.

MPTCP is the most well-known technique that allows automatic subpath failover
and load balancing to applications using TCP sockets.
We noted that the overhead is quite large so if a subpath is
very limited in throughput it is actually slower to aggregate
over the paths than to use both paths. So we developed a
LiFi monitor that uses a MPTCP path manager to configure
the subpaths accordingly if the LiFi (or Wi-Fi) connection gets to poor.

MPHTTP is very similar to MPTCP with the difference that only the client
needs to be modified to support the transmissions over multiple network interfaces.
The idea here is to make use of the HTTP Range query header and schedule 
ranges across Wi-Fi and LiFi based on the performance of each at any given time.

MPRTP is a new protocol we developed to allow WebRTC traffic to be load balanced
over LiFi and Wi-Fi and to instantaneously switch traffic from one or the other
without dropping any frames in a call.

Finally, we have also experimented with network priority on MacOS that allows you
to dynamically change which network interface should have highest priority.
When the LiFi antenna is directed towards you we can bump the LiFi interface
above the Wi-Fi interface to direct new applications to the LiFi network
without interrupting existing applications.

All these use cases actively probe to see if the LiFi network is usable with either
pings or various curl connection timeouts. That way they will be able to quickly take
advantage of a LiFi network that comes into range. Similarly they will be able to quickly
fail over to Wi-Fi when there is some temporary obstruction of LoS or the antenna is directed
away from the user.

The G.vlc LiFi specification implemented in chipsets today does not support seamless handover
with Wi-Fi, but the emerging 802.11bb specification defines handover support. When that specification is
implemented in chipsets, approaches like ours could benefit all applications seamlessly.

We note that our MASS synthetic workload generator is based on a GAN architecture which in turn relies on
a couple of DNNs, but since it its trained independently from 
the NN used to predict KPIs from workloads it does not taint the results.

In conclusion, we have seen in simulations and reproduced in experiments that a NN model 
is a promising predictor of negative impact as a basis for offloading decisions, and that the impact on network
KPI can be significant even when moving just a single user.

We also note that our approach is general enough to apply to other offload use cases. For example, one could imagine
using it to select the best client to steer to another band within a single technology too, such as across different Wi-Fi
spectrum regions.  

\section*{Acknowledgments}
It is a pleasure to acknowledge discussions with Bernardo A.~Huberman and Lili Hervieu of CableLabs.  We also thank Lili for a careful reading and critique of an earlier draft, and helping with the Wi-Fi experiment setup.

\bibliographystyle{IEEEtran}
\bibliography{related}

\pagebreak
\appendix
\section*{Appendix}
\subsection*{A Note on LiFi Antenna Alignment}\label{sec:alignment}
LiFi performance degrades both in terms of shortest path distance
from the antenna as well as horizontal displacement from the center of
the beam. Due to the latter better performance is achieved and longer
distances provide connectivity if the antenna is aligned with the receiver.
In the preceding experiment we pre-aligned the antenna to fixed positions
of different users. A more realistic scenario is where users may be located
at any number of positions in an area and we want to automatically align the
antenna quickly to where the user gets the best signal.

To this end we have developed an automatic alignment algorithm that
adjusts the antenna position to optimally align with the receptor.
We call the algorithm {\it pen tree} alignment as it is inspired by quad tree
search algorithms. The algorithm can be summarized as follows:
\begin{enumerate}
\item{In a rectangle defining the search area pick the corners and the center
as the five initial candidate positions.}
\item{From each candidate position measure the ping round trip times from the
user device to the LiFi antenna.}
\item{Create a Quartic Kernel Density Estimation (Q-KDE) heat map from the candidate points, and determine the heat
epicenter.}
\item{Divide each side of the original search area rectangle by half to produce
an area 1/4 of the original area.}
\item{Center the new area at the epicenter and recursively run the algorithm again
from the first step}
\item{The recursion stops as the area gets smaller and the epicenter does not change}
\end{enumerate}

We have experimented with this algorithm and our pan-tilt servo mechanism with promising
results for areas covering meeting room attendees in a 10 person meeting room, and
optical alignment could be achieved within a few seconds.

\end{document}